\documentclass[aps,prl,twocolumn,nofootinbib,floatfix]{revtex4} 

\usepackage{epsfig} 
 
\def\gsim{  \lower .75ex \hbox{$\sim$} \llap{\raise .27ex \hbox{$>$}} }  
\def\lsim{  \lower .75ex \hbox{$\sim$} \llap{\raise .27ex \hbox{$<$}} } 
\def\be{\begin{equation}} 
\def\ee{\end{equation}}

\def\mpl{M_{Pl}}

\usepackage{bm}
\usepackage{amssymb}
\usepackage{amsmath}

\def\fun#1#2{\lower3.6pt\vbox{\baselineskip0pt\lineskip.9pt
  \ialign{$\mathsurround=0pt#1\hfil##\hfil$\crcr#2\crcr\sim\crcr}}}

\def\fun#1#2{\lower3.6pt\vbox{\baselineskip0pt\lineskip.9pt
\ialign{$\mathsurround=0pt#1\hfil##\hfil$\crcr#2\crcr\sim\crcr}}}

\def\mpl{m_{\rm Pl}}
\begin{document} 

\title{Implications of a Running Spectral Index for Slow Roll Inflation}

 \author{Richard Easther}
 \affiliation{Department of Physics, Yale University, New Haven CT 06520, USA.}
\email{richard.easther@yale.edu}

\author{Hiranya V. Peiris\footnote{Hubble Fellow}}
\affiliation{Kavli Institute for Cosmological Physics and Enrico Fermi Institute,  University of Chicago, Chicago IL 60637, USA. }
\email{hiranya@cfcp.uchicago.edu}

\begin{abstract} We analyze the weak (2$\sigma$) evidence for a running spectral index seen in the three-year WMAP dataset and its implications for single field, slow  roll inflation. We assume that the running is comparable to the central value found from the WMAP data analysis,  and use the Hubble Slow Roll formalism to follow the evolution of the slow roll parameters. For all parameter choices consistent with a large, negative running, single field, slow roll inflation lasts less than  30 efolds after CMB scales leave the horizon. Thus, a definitive observation of a large negative running would imply that any inflationary phase  requires multiple fields or the breakdown of slow roll. Alternatively, if single field, slow roll inflation is sources the primordial fluctuations, we can expect the observed running to move much closer to zero as the CMB is measured more accurately at small angular scales. \end{abstract} 
\maketitle

\section{Introduction}

One of the most perplexing features of the 3-year dataset\footnote{We denote this WMAPII, and the one year dataset WMAPI.} from the Wilkinson Microwave Anisotropy Probe [WMAP]  \cite{Hinshaw:2006ia,Page:2006hz,Spergel:2006hy}  is that it hints at a significant ``running'' in the scalar spectral index.    This feature was seen, with  roughly the same level of significance, in the WMAPI analysis \cite{Spergel:2003cb,Bennett:2003bz}. The evidence for a running index persists when large scale structure data is added to the analysis \cite{Spergel:2006hy}, but is diluted by the addition of Lyman-$\alpha$ forest data, both with WMAPI \cite{Seljak:2004xh} and WMAPII \cite{Lewis:2006ma}. Joint analyses require a melange of datasets to increase their coverage in $k$-space and thus harbor the possibility of systematic normalization uncertainties or a tension in relative normalization between datasets, which could manifest itself as a spurious running. This source of error is eliminated when WMAPII is considered on its own, but only at the cost of reduced $k$-coverage.

We  stress that WMAPII does not demand a non-vanishing running, as zero is  $\sim2\sigma$ from the central value.  Rather, we spell out the consequences  of a large running -- particularly with a negative sign -- and the likely consequences for inflation. We find that a running similar to the WMAPII centroid rules out all simple models of inflation: those driven by a single, minimally coupled, scalar field whose evolution is well described by the slow roll formalism.  Inflation, if it happens at all, would thus be non-minimal. Alternatively, future analyses of datasets with a larger $k$-coverage will yield a running much closer to zero than the value extracted from WMAPII. 

We proceed by using the Hubble slow roll [HSR] hierarchy  \cite{hoffman/turner:2001,kinney:2002,easther/kinney:2003,liddle:2003,Peiris:2006ug}, which connects astrophysical determinations of the primordial power spectra (both tensor and scalar) to  the inflationary potential.  In \cite{Peiris:2006ug} we showed how a Monte Carlo Markov Chain [MCMC] analysis of  cosmological data can constrain the HSR parameters. These parameters are associated with flow equations that  determine their scale dependence.  We use this system of equations to compute the number of efolds of inflation which occur after a  mode  with wavenumber $k_0$ leaves the horizon.   Choosing $\xi$ large enough to explain the central value found for the running in Spergel {\em et al.\/} guarantees that any single field model of slow roll inflation will run for less than 30 e-folds after CMB scales leave the horizon.  As we explain below, one can increase the number of e-folds by adding a {\em fourth\/} slow roll parameter.  This term would be third order in the slow roll expansion,\footnote{The usual counting is that $\epsilon$ and $\eta$ are the lowest order terms, $\xi$ second order, and the next term is thus third order.} and the first three terms in the expansion would have roughly the same level of significance. Consequently,  the slow roll expansion cannot be safely used to  describe inflationary models with a substantial negative running. 

\section{The Slow Roll Approximation}

As summarized in  \cite{Peiris:2006ug}, the dynamics of single field inflation can be written in the Hamilton-Jacobi form, where overdots correspond to time derivatives and primes denote
derivatives with respect to $\phi$.   The HSR parameters $^{\ell}\lambda_H$ obey the infinite hierarchy of differential equations
\begin{equation}
\epsilon(\phi) \equiv \frac{m^2_{\rm Pl}}{4\pi}
\left[\frac{H'(\phi)}{H(\phi)}\right]^2; \label{eq:eps} 
\end{equation}
\begin{equation}
^{\ell}\lambda_H \equiv \left(\frac{m^2_{\rm Pl}}{4\pi}\right)^\ell
  \frac{(H')^{\ell-1}}{H^\ell} \frac{d^{(\ell+1)} H}{d\phi^{(\ell+1)}}
   ;\ \ell \geq 1.  \label{eq:hier}
\end{equation}
The usual slow roll parameters are $\eta = {}^{1}\lambda_H$ and $\xi =   {}^{2}\lambda_H$. If we set the higher order terms to zero at some fiducial point, these differential equations ensure they vanish at all other times.  The potential is given by
\begin{equation}
V(\phi) = \frac{3\mpl^2}{8\pi}    H^2\left(\phi\right)\left[1-\frac{1}{3}\epsilon\left(\phi\right)\right]  . \label{eq:v} 
\end{equation}
Liddle showed that the hierarchy can be solved exactly when truncated at order $M$ \cite{liddle:2003}, so  
\begin{equation}
\frac{H(\phi)}{H_0} = 1+ B_1 \left(\frac{\phi}{\mpl}\right) + \cdots + B_{M+1}
\left(\frac{\phi}{\mpl}\right)^{M+1} \, . \label{eq:h}
\end{equation}
The $B_i$ are specified by the initial values of the HSR parameters,
\begin{equation}
B_1 = \sqrt{4\pi\epsilon_0} \, \quad
B_{\ell+1} = \frac{(4\pi)^\ell }{(\ell+1)! 
\ B_1^{\ell-1}}  {}^{\ell}\lambda_{H,0}  \label{eq:coeffs}
\end{equation}
where the subscript $0$ refers to their values at the moment the fiducial mode $k_0$ leaves the horizon, and $\phi=\phi_0=0$. The number of e-folds, $N$ is given by 
\begin{equation}
\frac{dN}{d\phi} = \frac{4\pi}{\mpl^2} \frac{H}{H'} \, .  \label{dndphi}
\end{equation}
 Finally,  $\phi$ and $k$ are related by
\begin{equation}
\frac{d\phi}{d\ln k} = -\frac{\mpl}{2\sqrt{\pi}}
\frac{\sqrt{\epsilon}}{1-\epsilon}. \label{eq:phieq}
\end{equation}

We now turn to the inflationary observables, 
\begin{eqnarray}
n_s &=& 1+ 2\eta - 4\epsilon - 2(1+{\cal C}) \epsilon^2 - \nonumber \\
  && \quad \frac{1}{2}(3-5{\cal C}) \epsilon \eta + \frac{1}{2}(3-{\cal C})\xi \\  \label{eq:hsrconversionsstart}
r &=& 16 \epsilon \left[1+2{\cal C}(\epsilon - \eta)\right]\\
 \alpha &=& -\frac{1}{1-\epsilon} \left\{2 \xi - 8 \epsilon^2 - 10 \epsilon \eta +  \frac{7{\cal C}-9}{2} \epsilon \xi  \right. \nonumber \\
&& \quad \left. + \frac{3-{\cal C}}{2} \xi \eta \right\}  \label{eq:nsrun}
\end{eqnarray}
where ${\cal C}=4(\ln 2 + \gamma) - 5$, and we have introduced the customary notation $\alpha  = dn_s/ d\ln{k}$ and $r$ is tensor:scalar ratio.   We retain all terms in $\alpha$ up to quadratic order in the slow roll parameters, anticipating that $\xi$ may be as large as $\epsilon$ or $\eta$.

\begin{figure}[tb]
\includegraphics[height=2.7in,width=3.3in]{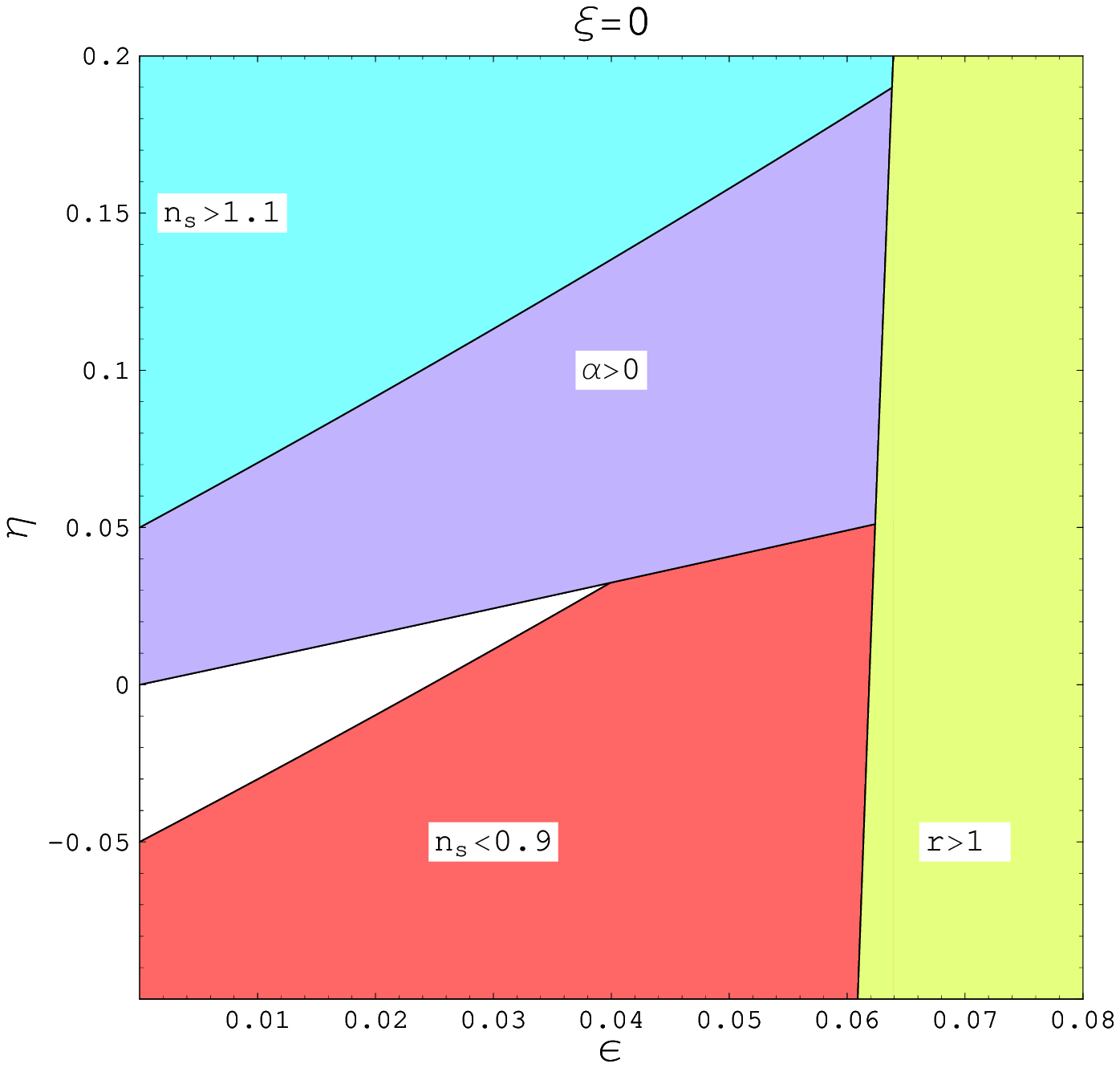}
\includegraphics[height=2.7in,width=3.3in]{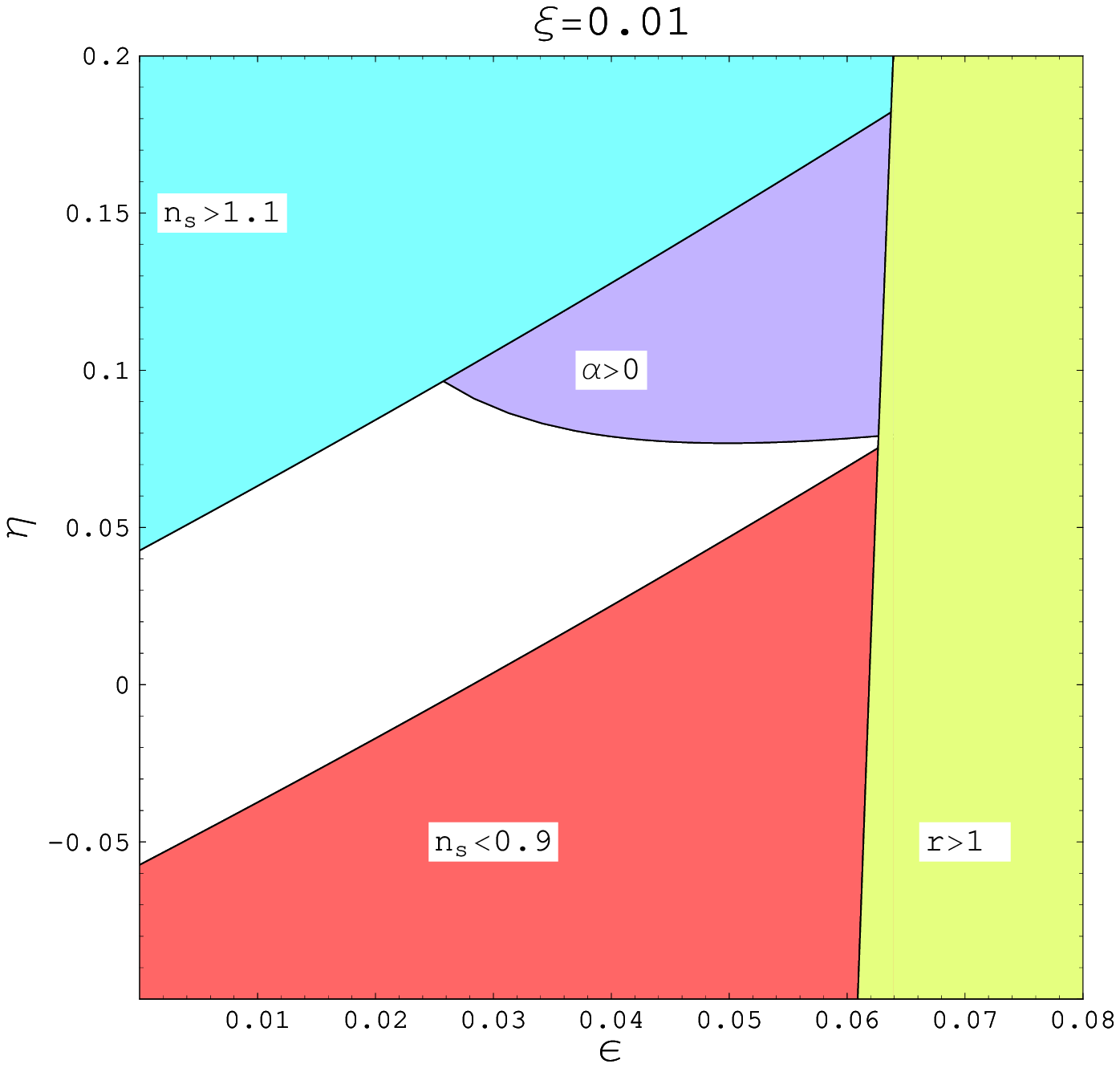}
\caption{We show the regions of the $(\epsilon,\eta)$ plane excluded by the assumed bounds on the spectral parameters. As $\xi$ increases, the portion of the plane for which $\alpha>0$ shrinks. \label{pl:cuts1}}
\end{figure}

\begin{figure*}[tb]
\includegraphics[height=2.7in,width=3.3in]{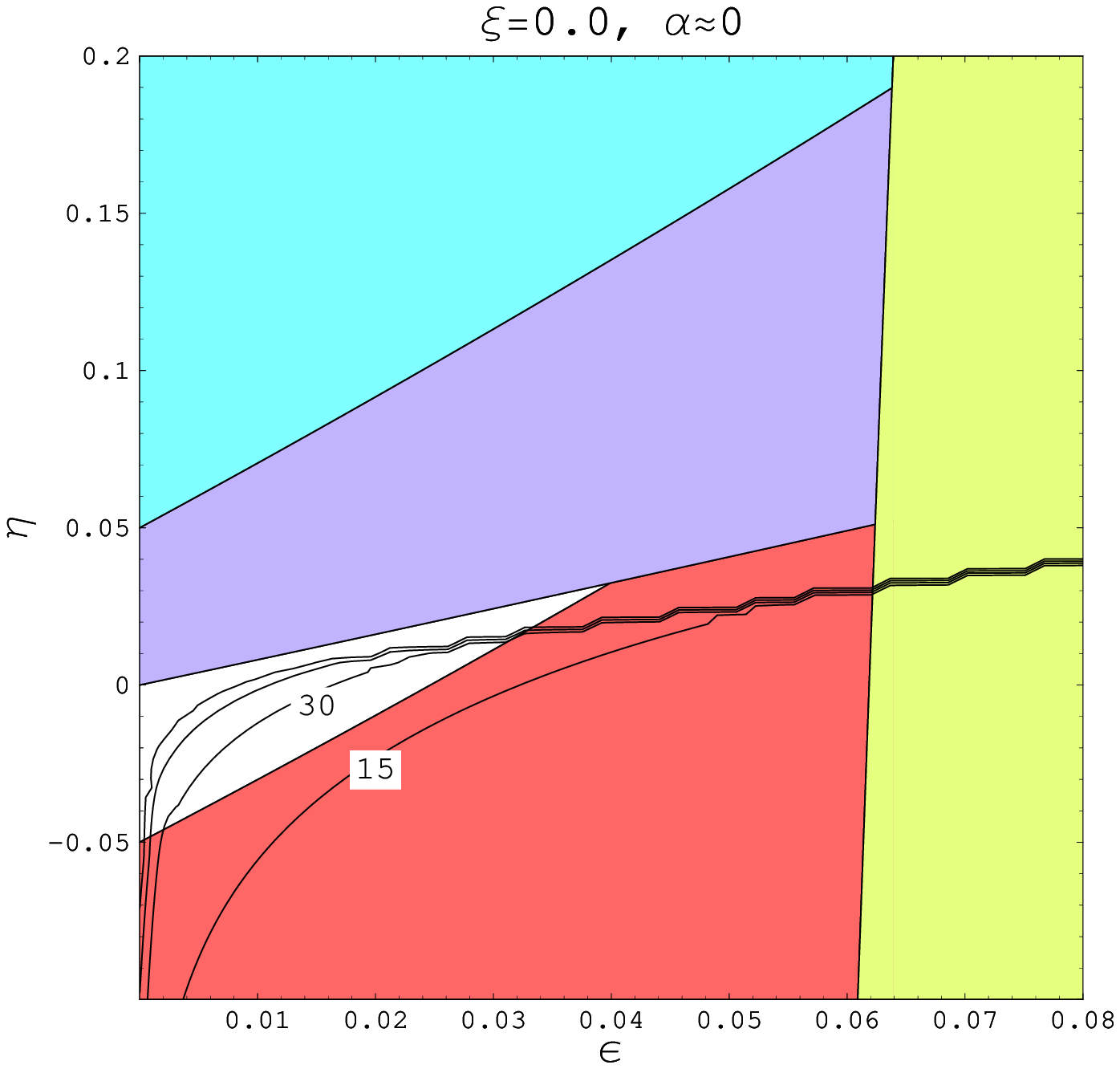}
\includegraphics[height=2.7in,width=3.3in]{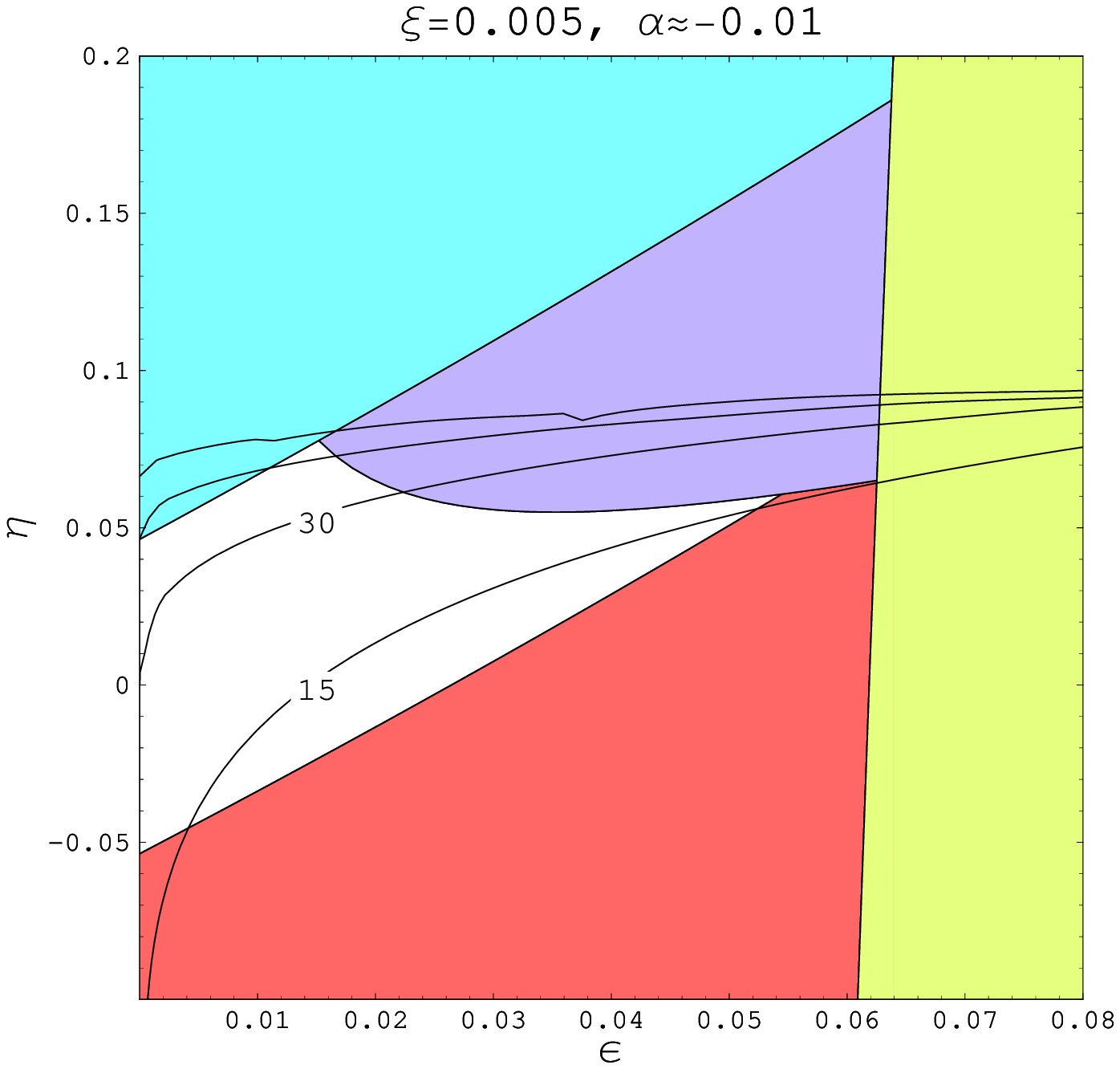}
\includegraphics[height=2.7in,width=3.3in]{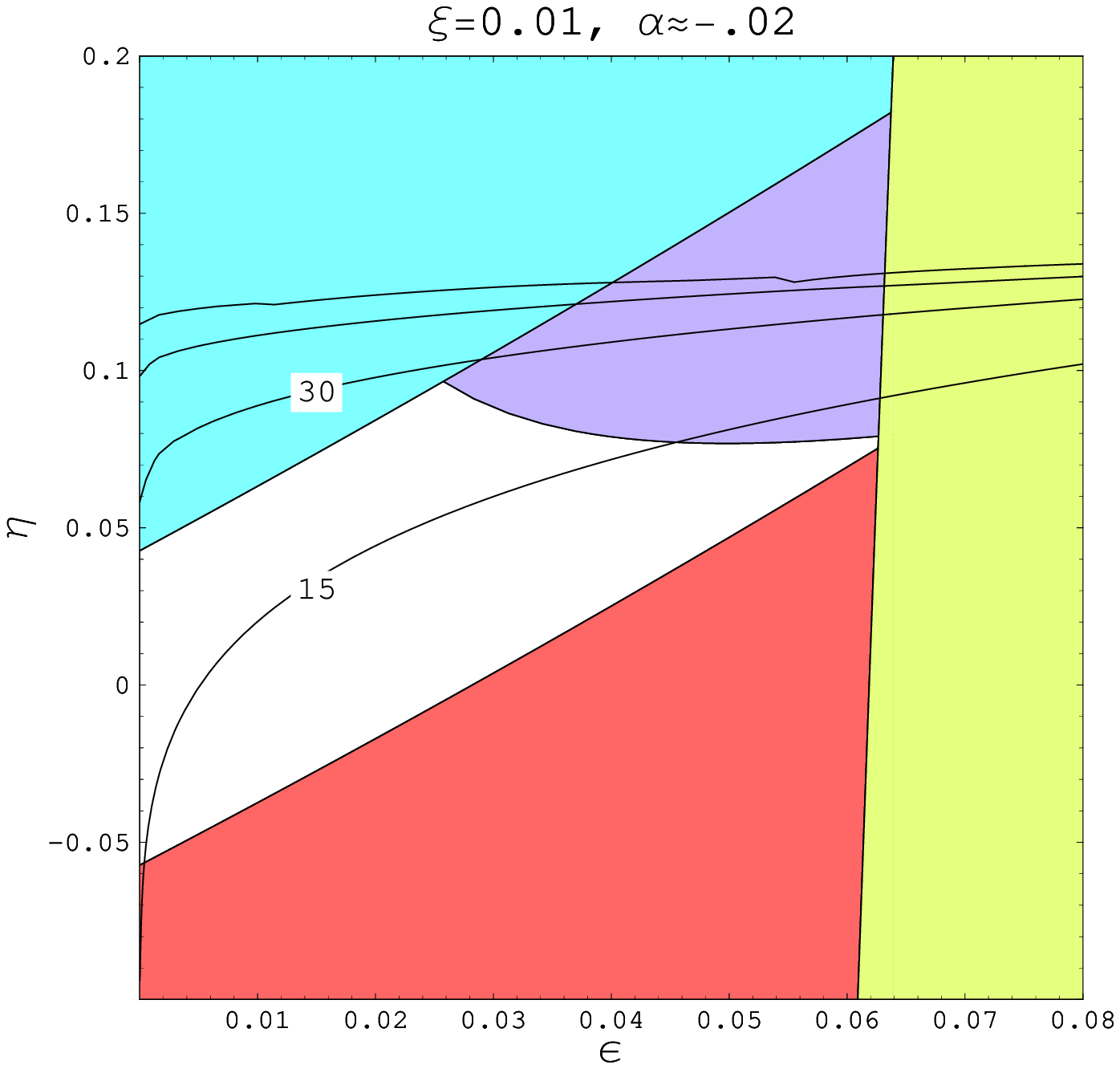}
\includegraphics[height=2.7in,width=3.3in]{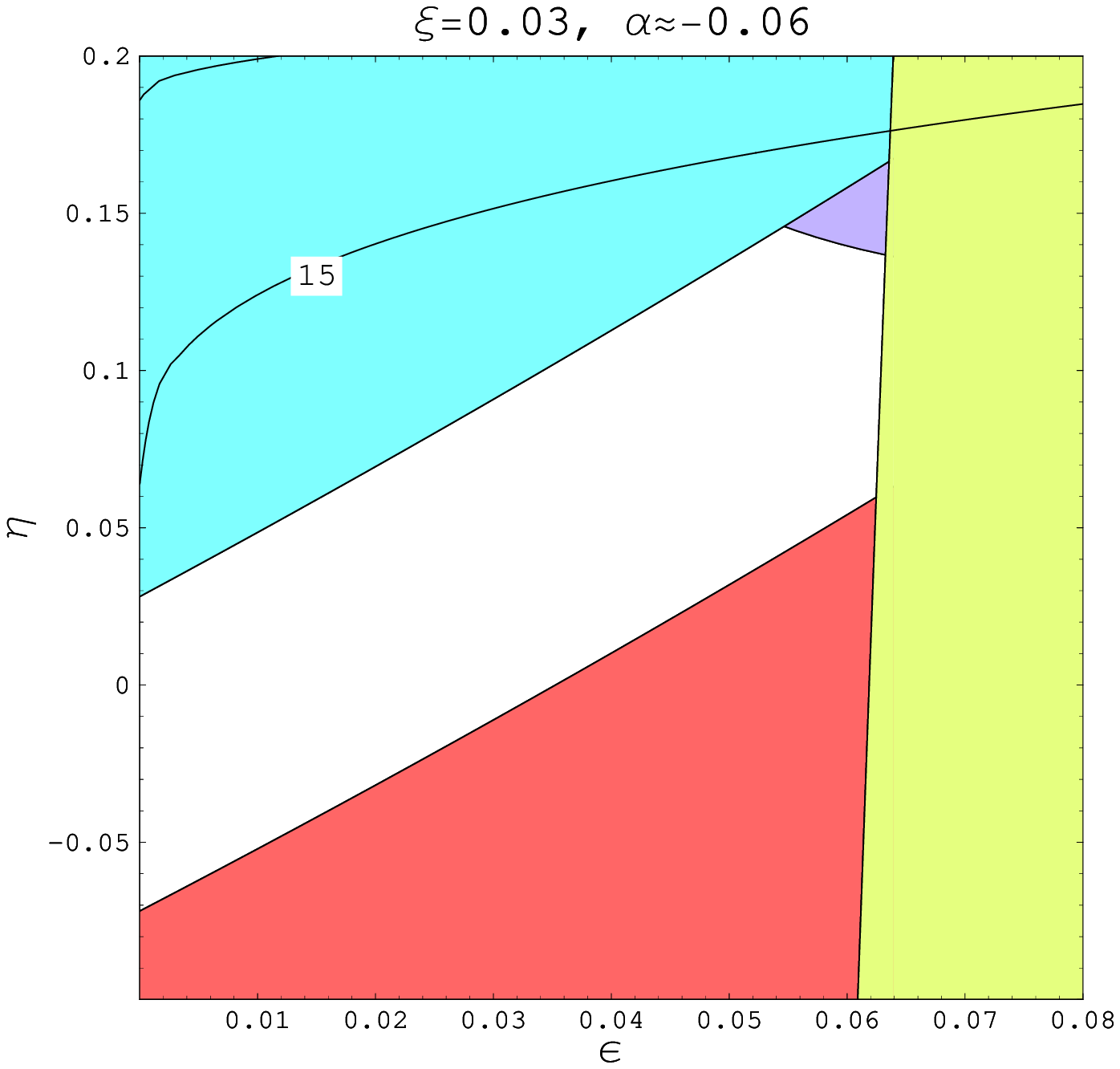}
\caption{We show the regions of the $(\epsilon,\eta)$ plane excluded by the assumed bounds on the spectral parameters for the same cuts displayed in Figure~\ref{pl:cuts1}. We include contours (for $N=15,30,45,60$) showing the number of e-foldings occuring after the fiducial mode $k_0 =0.002\mbox{Mpc}^{-1}$ leaves the horizon. For $\xi =0.03$ the only regions of the $(\epsilon,\eta)$ plane with an acceptable value of $n_s$ have $N<15$. The running is a weak function of $\epsilon$ and $\eta$ -- the values of $\alpha$ above are taken from $\xi$ alone. \label{pl:cuts2}}
\end{figure*}

\section{Cosmological Constraints}

Consider the following very weak constraints 
\begin{eqnarray}
&0.9 < n_s < 1.1& \label{eq:ncut} \\ 
&r<1 .& \label{eq:rcut}
\end{eqnarray}
These bounds are broad enough to be independent of any assumptions about the presence or absence of a running index, and are outside the 99\% confidence intervals derived from WMAPII and other datasets \cite{Spergel:2006hy}.    The running is given by equation \ref{eq:nsrun} -- $\xi$ is the only term linear in the slow roll parameters and it dominates the expression when $|\alpha|$ is large.  In particular, for large negative $\alpha$, we need $\xi>0$. While $\epsilon$ and $\eta$ can become large for fixed $n_s$, they make a positive contribution to $\alpha$ thanks to the coefficients on $\epsilon\eta$ and $\epsilon^2$ in equation \ref{eq:nsrun}.   Figure~\ref{pl:cuts1} shows the cuts these constraints put on the $(\epsilon,\eta)$ plane for $\xi =0$ and $\xi =0.01$.  Finally,  we show the region where $\alpha > 0$.   

Again we remind the reader that WMAPII does not require that $\alpha <0$ and the purpose of this Letter is to explore the consequences of measuring  a value of $\alpha$ near the WMAP centroid.  This number depends on whether one allows for a contribution from primordial tensors. For the pure scalar case, the distribution peaks at $\alpha \sim -0.05$. Including tensors actually makes the central value larger, but this is not necessary for us to establish the conclusions we reach below.

The minimal required amount of inflation is not well defined. If inflation is to happen before the electroweak phase transition, the number of e-folds $N$ must be greater than 30, and $N\sim 55$ for GUT scale inflation. These numbers are mildly dependent on the reheating mechanism, but are sufficient for our purposes.  Any model with $N<30$ is unlikely to provide a workable explanation for the large scale homogeneity and isotropy of the universe.    Because we can compute the running of the HSR parameters  as a function of $\phi$ (or, equivalently, $k$), we can obtain the remaining number of e-folds for any choice of $\{\epsilon_0,\eta_0,\xi_0\}$.  The end of inflation is signified by the instant when $\epsilon =1$, and we find $N$ from  equation (\ref{dndphi}) alongside the flow equations, (\ref{eq:eps}) and (\ref{eq:hier}). The result of this calculation is displayed in Figure~\ref{pl:cuts2} for four values of $\xi$.  Truncating the slow roll hierarchy at third order and choosing $\xi$ consistent with a large negative running, we see that the fraction of  parameter space that has $N>30$ and $\alpha \lesssim -0.02$ is of measure zero.

\section{Discussion}

In the above analysis we have shown that a large, negative of $\alpha$, the running  of the scalar spectral index, cannot be produced by a self-consistent single field inflationary model which is well described by the slow roll expansion.  Specifically, the large running leads to  an unacceptably small value of $N$, the number of e-folds. The low value of $N$ can be ameliorated by adding a further slow roll parameter ($^3\lambda$) to the analysis which, if carefully chosen, can move $\xi(\phi)$ closer to zero as inflation continues.   However, in this case the inflationary dynamics would be described by four numbers, $\{\epsilon_0,\eta_0,\xi_0,^3\lambda_0\}$, all of roughly equal importance.   Such a model may be ``slowly rolling'' in the sense that $\epsilon \ll 1$, but the slow roll {\em expansion\/} is not  trustworthy when its first three orders are of roughly equal weight. Nor is there is any guarantee it could be safely truncated after $^3\lambda_0$.

If the central value of $\alpha$ derived from the WMAPII analysis (with or without large scale structure information) is near the actual value, all models of inflation driven by a single, minimally coupled, slowly rolling scalar field are ruled out, and a large class of inflationary models would thus be falsified.   There is no guarantee that inflation is well described by slow roll. Models with ``features'' in the potential yielding a local violation of the slow roll conditions have been considered in the past (e.g. \cite{Starobinsky:1992ts,Adams:2001vc,Peiris:2003ff}).   Further, models of inflation with two or more interacting fields have much more freedom to yield complicated spectra \cite{Sasaki:1995aw}, or one could turn to scenarios with two or more bursts of inflation \cite{Silk:1986vc,Holman:1991ht,Polarski:1992dq,Lyth:1995ka,Burgess:2005sb}.  However, invoking any   of these options would dramatically complicate the theoretical understanding of inflation. 

This analysis depends on the HSR equations, and their ability to incorporate constraints arising from the duration of inflation alongside those from the perturbation spectra  \cite{Peiris:2006ug}.     The role of the $\xi$ (or ``jerk'') term in the dynamics of the flow equations is discussed in detail in \cite{Chongchitnan:2005pf}. This analysis focuses on ``attractors'' in the slow roll parameter space, in the presence of non-zero high order terms in the slow roll hierarchy, rather than their end-state, which is our principal concern here.   Likewise, inflationary models generated via the flow equations \cite{kinney:2002,Peiris:2003ff} or  {\em Monte Carlo Reconstruction\/} \cite{easther/kinney:2003} can provide solutions with a large, negative running and $N \gsim 55$, but these  rely on contributions from higher order terms.  The accuracy of a specific class of slowly rolling models with significant running is probed using Monte Carlo Reconstruction techniques in \cite{Makarov:2005uh}. Conversely, \cite{Cline:2006db} argues that the support for running in the  WMAPI dataset comes from the multipoles with $30 \lesssim \ell \lesssim 40$.    Finally, \cite{Malquarti:2003ia} also point to the tension between a large running and providing the required number of e-folds.   However, our conclusion: that no single field  model  can simultaneously a) satisfy slow roll, b) produce a large negative running and c) provide a sufficient number of efolds, is new.

Of course,  by far the most prosaic  resolution of this conundrum would be that the weak evidence for a significant running in the WMAPII dataset is a statisical artifact that will evaporate as more data becomes available. Indeed, adding Lyman-$\alpha$ forest data is known to reduce the need for a negative value of $\alpha$ \cite{Seljak:2004xh,Lewis:2006ma}.   In particular, we anticipate that the Planck mission will significantly  enhance our understanding of $\alpha$ by providing high quality measurements of the fundamental power spectrum over a larger wavelength range than WMAP, without the need to combine several heterogeneous datasets that carry the risk of systematic biases in their relative normalization.

\section*{Acknowledgments } We   are grateful to  Will Kinney,  Eugene Lim, and members of the WMAP science team for useful discussions.  RE is supported in part by the United States Department of Energy, grant DE-FG02-92ER-40704.  HVP is supported by NASA through Hubble Fellowship grant \#HF-01177.01-A awarded by the Space Telescope Science Institute, which is operated by the Association of Universities for Research in Astronomy, Inc., for NASA, under contract NAS 5-26555.

\end{document}